\documentclass[aps,prl,10pt,showpacs,reprint,superscriptaddress,floatfix]{revtex4-1}
\usepackage[space]{grffile}
\usepackage{latexsym}
\usepackage{textcomp}
\usepackage{longtable}\usepackage{multirow,booktabs}
\usepackage{amsfonts,amsmath,amssymb}

\usepackage{url} \usepackage{hyperref}
\hypersetup{colorlinks=false,pdfborder={0 0 0}}

\usepackage{graphicx,color,epsfig}
\usepackage{colortbl} \newcolumntype{C}{c<{\kern\tabcolsep}@{}}
\usepackage[table]{xcolor} \usepackage{stfloats}
\usepackage[utf8]{inputenc}

\definecolor{Gray}{gray}{0.9} \definecolor{LGray}{gray}{0.7}
\definecolor{Blue}{RGB}{175,238,238}
\definecolor{LBlue}{RGB}{224,255,255}
\begin{document}

\title{The web of federal crimes in Brazil: topology, weaknesses, and control}
\author{Bruno Requião da Cunha}\email{cunha.brc@dpf.gov.br}
\affiliation{Polícia Federal,Brazil}
\affiliation{Instituto de Física, Universidade
  Federal do Rio Grande do Sul, Porto Alegre, RS, Brazil}
\author{Sebastián Gonçalves}\email{sgonc@if.ufrgs.br}
\affiliation{Instituto de Física, Universidade
  Federal do Rio Grande do Sul, Porto Alegre, RS, Brazil}

\date{\today}

\begin{abstract}
Law enforcement and intelligence agencies worldwide struggle to find
effective ways to fight and control organized crime.  However, illegal
networks operate outside the law and much of the data collected is
classified. Therefore, little is known about criminal networks
structure, topological weaknesses, and control.  In this contribution
we present a unique criminal network of federal crimes in Brazil. We
study its structure, its response to different attack strategies, and
its controllability.  Surprisingly, the network composed of multiple
crimes of federal jurisdiction has a giant component, enclosing more
than a half of all its edges. This component shows
some typical social network characteristics, such as small-worldness
and high clustering coefficient; however it is much ``darker'' than
common social networks, having low levels of edge density and network
efficiency.  On the other side, it has a very high modularity value,
$Q=0.96$.  Comparing multiple attack strategies, we show that it is
possible to disrupt the giant component of the network by removing
only $2\%$ of its edges or nodes, according to a module-based
prescription, precisely due to its high modularity.  Finally, we show
that the component is controllable, in the sense of the
exact network control theory, by getting access to $20\%$ of the
driver nodes. 
\end{abstract}

\maketitle

\section{Introduction}
Despite recent efforts of Brazilian law enforcement agencies in
combating organized crime, the horizon looks no promising: homicide
rates have spiked in 2014 reaching 29.1 deaths per hundred thousand
people~\cite{Cerqueira:2016aa}, the country has become the second
greatest consumer of cocaine in the world ---turning into one of
  the most important corridors for international drug trafficking---,
corruption and money laundry have pervaded major enterprises and
important political figures nationwide~\cite{UN2015}.

The problem is multivariable, from cultural and historical issues to
the structure of the judicial and law enforcement systems. However,
from the network science point of view, another important reason lies
on the structure of the web of crime and on the characteristics of
traditional police interventions.
Law enforcement actions consist mainly in random arrests and
operations focusing on big criminal leaders, in a strategy loosely
resembling a degree-based attack.  Nonetheless, it was shown that the
social structure of criminal activities is, in fact, highly resilient
to the traditional law enforcement approach~\cite{Morselli:2009aa,
  Spapens:2011aa, Morselli:2007ab}.
In this sense, as pointed out in earlier studies, there is a lack of
directions or focused strategy in crime fighting which, most of the
times, is reactive, always standing one step behind criminal
undertakings~\cite{Natarajan:2006aa, Sarnecki:2001aa, Chen:2004aa,
  Drezewski:2015aa}. Thence, the necessity to address law enforcement
in a proactive framework and crime as a collective complex system
brings us to the study of this criminal network.

The sociological literature vastly supports both theoretically and
experimentally the adoption of network methods in studying criminal
rings~\cite{McGloin:2005aa, Sah:1991aa, Glaeser:1996aa,
  Morselli:2003aa}. For instance, the social facilitation
model~\cite{Mastrobuoni:2012aa, Thornberry:1993aa} states that the
impulse of an individual to criminal actions is somehow strengthened
by his or hers membership to a criminal organization. Therefore,
atomizing the network structure of a criminal organization would
lower, theoretically, the crime rates. In this sense, many studies in
social science have focused in efficient ways to dismantle this
dreadful phenomenon. However, these contributions are mostly based on
qualitative and loose aspects of individuals rather than on their
quantitative and collective role in maintaining the network
functioning as a whole (\textit{i.e.} their topological
centralities)~\cite{Ballester:2006aa, Borgatti:2006aa}.  
Several researchers have recently
illustrated the benefits of applying network science and statistical
physics~\cite{DOrsogna:2015aa} to study the structure and fragility of
the criminal phenomenon. For instance, Agreste \textit{et al.} have
recently studied the network structure and resilience of the Sicilian
Mafia (often known as \textit{Cosa Nostra})~\cite{Agreste201630}. In
that paper, the cooperation with Italian law enforcement agencies
yield to a bipartite network (contact and criminal), which showed
different robustness to network attacks ---the contact network is much
more fragile to targeted attacks than the criminal one. Despite that,
the authors did not study the Mafia network's modularity and its
robustness to Module-Based Attacks(MBA)~\cite{Requiao-da-Cunha:2015aa}
or other more efficient methods of attack~\cite{Morone:2015aa}.
Besides, other authors have studied Mafia syndicates, pointing to the
strong hierarchical networked organization with a few \textit{cappos}
(bosses) commanding the criminal activities~\cite{Cayli:2013aa}.
\begin{figure}
\begin{center}
\includegraphics[width=0.95\columnwidth]{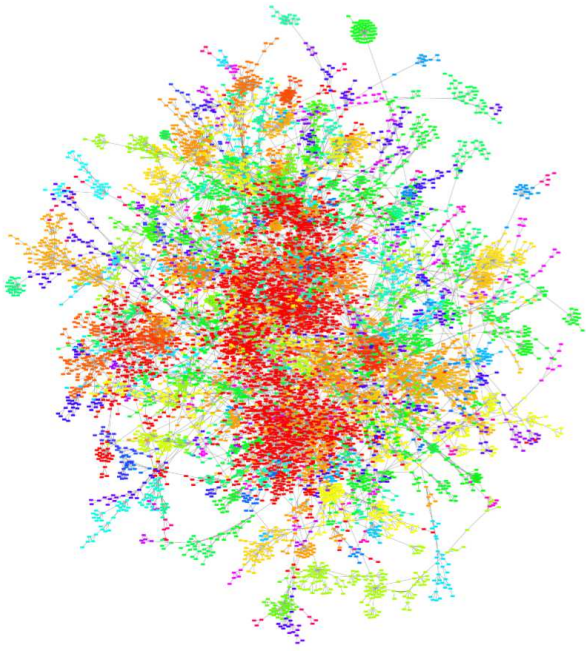}
\caption{Representation of the largest connected component of the
  federal crimes network consisting of 9,887 individuals and 91
  modules. Colors represent nodes from the same communities as
  extracted by the Louvain method.
  \label{palas}}
\end{center}
\end{figure}  

Due to the economic-driven nature of organized crime, a common feature
of criminal systems is the balance between the secrecy of its illegal
activities and the efficiency of its communications and operations. As
shown in earlier researches, these traits are directly related to the
network structure of the criminal phenomenon~\cite{Toth:2013aa}.
Therefore, such systems need stable action-based relationships which
increases network efficiency and consequently results in higher risks
for the illegal enterprise. Precisely because of that, criminal
networks tend to be fragile to targeted
attacks~\cite{Morselli:2007aa}.  However, Duijn \textit{el
  al.}~\cite{Duijn2014} pointed out recently, when studying a drug
related network from the Dutch Police, that criminal organizations may
become more efficient as a response to targeted attacks. The positive
counterpart is the decrease of the network's security, offering
strategic opportunities for law enforcement and intelligence agencies
to plan effective network disruption operations.

In order to address precisely this phenomenon, as part of an ongoing
collaboration with Brazilian Federal Police, we had access to data
collected from several federal crimes which result in a web of almost
24,000 individuals across several layers of criminal relationships.
This unique set of data allows us to study the social networks
adjacent to the criminal phenomenon and its topological weaknesses.
Therefore, it is of high practical importance to search for the most
efficient way to shatter these social networks in order to help
reducing criminal levels


Consequently, in this contribution we aim at describing the structure
of such network, finding effective methods to disrupt it, and to control
the activities on such a system.

The paper is organized as follows: we first present the collected
data, next we describe the underlying topology of the resulting
network; then, we study the network's fragility to targeted attacks;
after that, we explore the controllability of the network. At the end,
the main results are summarized and conclusions are drawn.

\section{Dataset and method}  
The dataset was obtained by the office of the Brazilian Federal
Police at Rio Grande do Sul and anonymized before releasing for
scientific, academic, and collaborative purposes. The record of
criminal investigations were gathered from April 2013 to August 2013 and
consist of information provided directly by each case's investigator
who would inform, through a computer application (\emph{i.e.} a
digitized version of card files), the
proper relationship among the individuals investigated by he or she
resulting in $N=23,666$ individuals (nodes) and $E=35,913$
relationships (edges).
The query consisted of investigations
concerning all federal crimes, a set of very different crimes such as
organized crime, money laundering, international drug trafficking,
terrorism, international pedophilia rings and corruption schemes. The
original police database of criminal records contains classified
information. On account of that, data was filtered and anonymized in
order to protect individual rights and comply with legal
requirements. Topological features of the relationships were preserved
in order to study the adjacent network structure. The anonymized
network data is available at Konect (http://konect.uni-koblenz.de/).

The resulting undirected and unweighted network has $3,425$
unconnected components with a medium size of only $7$
individuals. However, the degree dispersion $\langle
k^{2}\rangle/\langle k\rangle=7.42$ is much higher than the
Molloy-Reed criterion meaning the network is actually in a regime
which there is a giant component pervading the whole
system~\cite{Dorogovtsev:2013aa}. This is the first important result
of our study, for it was not expected that a giant component would
rise in a set of actors committing criminal actions not related in
principle to one another-- such as drug trafficking and
pedophilia. Therefore, from now on we focus only on the giant
component of the system, since the fragmented pieces might be
considered as a residual criminal phenomenon characteristic of every
society while the largest connected component represents a generalized
and self-organized criminal phase that is more dangerous from a
national security point of view that should be concerned carefully by
federal and national law enforcement and intelligence agencies. The
largest connected component consists of $9,887$ nodes and $19,744$
edges ($40\%$ of the total number of nodes and $54\%$ of the total
number of edges, see Fig.~\ref{palas})

\begin{table}  
\centering
\begin{tabular}{ r | r r r r}
  & N  &  E  & $\delta$ & $\eta$\\
\hline
Facebook (NIPS) & 2888 & 2981 & 0.0071 & $29.3\%$ \\
Hamsterster & 2426 & 16631 & 0.0056 & $20.8\%$ \\
Crime & 829 & 1473 & 0.0043 & $21.5\%$ \\
PF (Fig.~\ref{palas}) & 9887 & 19744 & 0.0004 & $8.4\%$ \\
\end{tabular}  
\caption{Comparative data between the federal crimes network and other
  social networks: Number of nodes ($N$), number of edges ($E$), edge
  density ($\delta$), and graph efficiency ($\eta$)
  for four distinct social networks:
  a Facebook user-user friendship network~\cite{konect2014egofacebook,konectMcAuley2012},
  a friendship network for the user of the website hamsterster.com~\cite{konect:2016:petster-hamster},
  a criminal dataset recorded by St Louis Police in the 1990s~\cite{konect:2016:moreno_crime},
  and the federal crimes network of this contribution.
  \label{tabela1}}
\end{table}

\section{Network structure}
\begin{figure}
\begin{center}
\includegraphics[width=0.95\columnwidth]{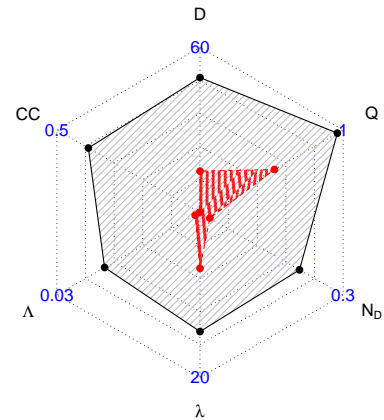}
\caption{The radar chart displays the following network parameters for
  the federal crimes data (gray pattern)and its randomized counterpart
  (red pattern): diameter ($D=49$ and $15$), average shortest path
  length ($\lambda=14.43$ and $6.78$), assortativity ($\Lambda=0.017$
  and $0.001$), clustering coefficient ($CC=0.391$ and $0.001$) and
  modularity ($Q=0.96$ and $0.52$).\label{radar}}
\end{center}
\end{figure}

The density and the efficiency of a criminal network is usually
related to the ``brightness'' of the system in the sense that a large
number of connections among criminals means that if one actor is
caught by law enforcement or intelligence agencies it would be
possible, at first, to extract critical information about the
network's structure~\cite{Duijn2014}.  On the other hand, a darker
network means the direct transfer of information within the system is
slowed down due to the decreased number of possible path among
criminals. Therefore, both the network density and the network
efficiency informs us about the compromise between security and
effective diffusion of information and data. This is precisely the
case of our network which is ``darker'' than some traditional social
networks, \textit{i.e.} it has lower edge density levels, and at the
same time it has a very low graph efficiency (see
Table~\ref{tabela1}).  The radar chart presented in Fig.~\ref{radar}
shows the topological differences between the criminal network and its
randomized version, where all edges are rewired. The data highlight
the small-worldness~\cite{Dorogovtsev:2013aa} of the system since the
network has small average shortest path length as compared to the
network's diameter but with a larger clustering coefficient.

\begin{figure}
\begin{center}
\includegraphics[width=0.95\columnwidth]{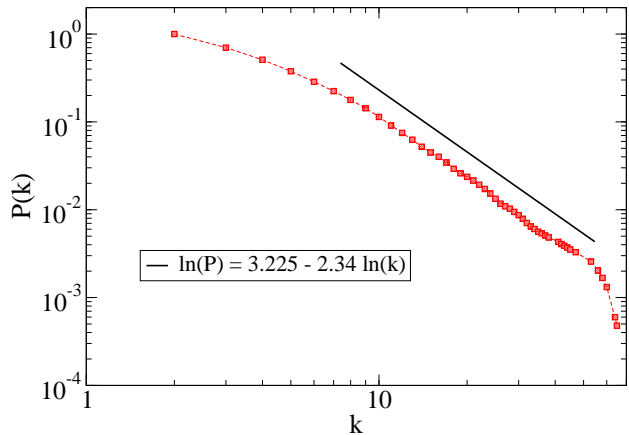}
\caption{Cumulative degree distribution for the federal crimes network
  with log-log axis and a power law fit in the middle region
  ($6 < k < 53$): $P(k)\sim k^{-\gamma}$
  with $\gamma\simeq 2.34$.\label{cum_deg}}
\end{center}
\end{figure}

The degree distribution of a graph is of utmost importance in
unraveling the nature of its adjacent system. For instance, besides
other practical implications, networks with homogeneous degree
distributions, which the probability $p(k)$ that an arbitrary node has
degree $k$ decays exponentially for large values of $k$, face a
transition from a fully connected to a disconnected phase if a
fraction $q_{c}$ is randomly removed from it~\cite{netscibook}. On the
other hand, graphs in which $p(k)$ has a heterogeneous distribution
are usually robust to random failure of nodes but weak to targeted
attacks to its most central nodes or hubs~\cite{netscibook}. Examples
of heterogeneous systems include the Internet, the World Wide Web, and
in general most (large-scale) social
systems~\cite{Dorogovtsev:2013aa}. In this sense, degree distributions
that follow a power-law ($p_{i}\propto k_{i}^{-\gamma}$, where $p_{i}$
is the probability a node attaches to $i$, which has already a degree
$k_{i}$) are usually known as scale-free and reveal a generative model
with preferential attachment in which nodes tend to connect to the
more popular nodes in a rich-gets-richer mechanism.  However, in real
systems the degree might not be the only factor which attracts
connections, in social networks for instance several qualities might
interest reciprocal relationships such the sharing of same political
or ethical vision.  Such attributes are therefore called the vertex's
fitness and might generally be expressed by hybrid multiplicative
processes such as the Log-normal Fitness Attachment
(LNFA)~\cite{Ghadge:2010aa}
\begin{equation}
p_{i}(k)\propto k_{i}\prod_{l}\phi_{il}
\end{equation}
where $\phi$ represents the set of attributes of node $i$. When the
number of attributes are sufficiently large and statistically
independent, it is shown that the fitness is log-normally distributed,
regardless of the type of the particular distribution of each
attribute~\cite{Nguyen:2012aa, Bell:2017aa}.  In the same way, the
degree distribution of the federal crime network has an approximate
log-normal behavior, with a power-law regime in the middle of the
cumulative representation showing an exponent $\gamma \simeq 2.34$ (see
Fig.~\ref{cum_deg}), which is very similar to the Mafia network studied
by Agreste \textit{et al.}~\cite{Agreste201630} which shows a power-law
regime with $\gamma \simeq 2.5$
This means that in this criminal system there is an interplay between
at least two distinct phenomena: the tendency of individuals to
connect not only to hubs, but also to fittest criminals. This
particular multiplicative statistic can be explained by group
affinities such as gangs, mafia and mobs which also explains the
modular architecture of the network as it is shown later in the
text--- Brazil has a long history of well defined and competing
criminal syndicates such as the \emph{Primeiro Comando da Capital} (First
Command of the Capital or PCC) and the \emph{Comando Vermelho} (Red Command
or CV) among others, summing up more than 10,000 individuals
nationwide~\cite{McCann:2007aa}.  That is, among individuals with
similar number of connections (degree), new criminals will connect
most likely to the ones belonging to the criminal gang with stronger
influence in their social medium (higher fitness).  Conversely, if the
fitnesses are similar, the vertex with highest degree will more
probably be selected.


\section{Network disruption \& controllability}  
From a network science point of view, a graph can be refrained from
functioning as a whole either by removing its nodes or by removing
only its edges (maintaining the nodes). In this sense, law enforcement
operations usually aim at identifying and arresting
criminals. Therefore, the arrest of individuals is directly related to
the removal of edges, since the nodes are not in fact deleted from the
network. On the other hand, the deletion of nodes means the complete
removal of that individual from the criminal network-- a scenario that
would only occur either by death or by re-socialization and not
directly by law enforcement actions. In a topological perspective,
node removal is more effective in atomizing complex networks causing
more damage per elimination than edge removal since the deletion of a
single node from the network results in the elimination of all links
attached to it ~\cite{Iyer2013,Crucitti2004}.  This is a second
important result, with important sociological implications,
\textit{ie} from a network science perspective, re-socialization
(\emph{eg} by education or by work) is in general a more effective
strategy to lower crime levels than imprisonment. Still, it should be
noted that according to this rationale, even though reprehensible
ethically and legally, the death of the key individuals would reach
similar results, \emph{ceteris paribus}.

\begin{figure*}
\begin{center}
\includegraphics[width=\linewidth]{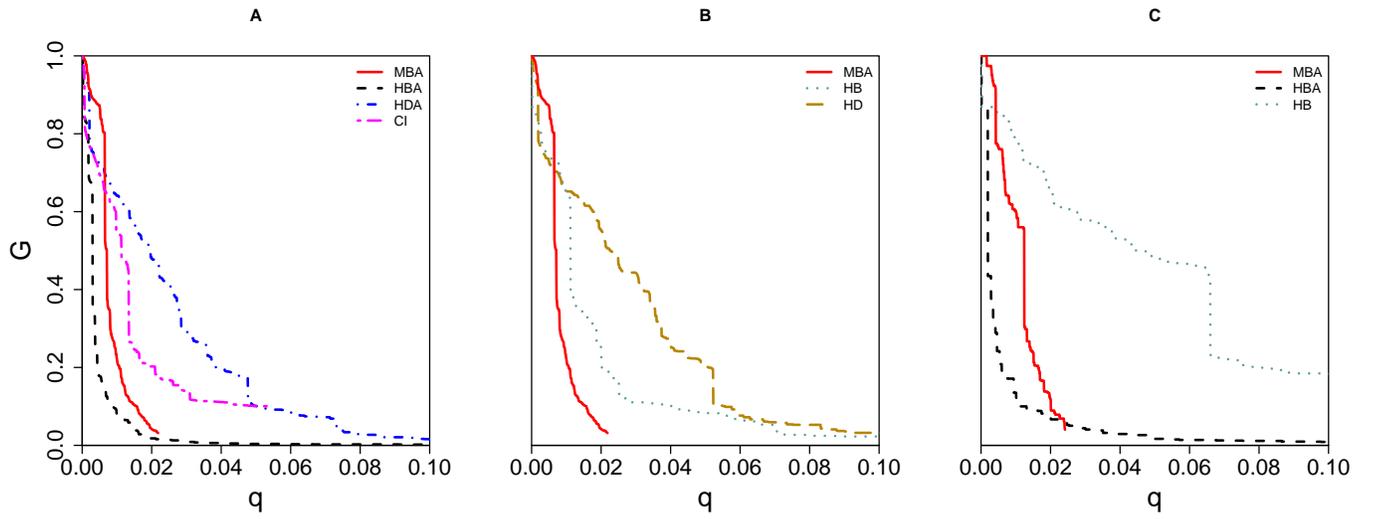}
\caption{The figures show the fragmentation curves of the federal
  crimes network by the relative size of the largest connected
  component $G$ as a function of the fraction of nodes or edges
  removed $q$, according to different procedures: (A), node-based High
  Degree Adaptive (HDA - blue dot dashed line), High Betweenness
  Adaptive (HBA - black dashed line), Module-Based attacks (MBA - red
  solid line), and Collective Influence (CI - magenta two dashed
  line); (B), node-based High Betweenness (HB - light blue dotted
  line), High Degree (HD - golden long dashed line), and Module-Based
  (MBA - red solid line) attacks; (C), edge-based High Betweenness
  Adaptive (HBA - black dashed line), High Betweenness (HB - light
  blue dotted line), and Module-Based attacks (MBA - red solid
  line).\label{sigmas}}
\end{center}
\end{figure*}

We now simulate both edge and node disruptions to the network's giant
component by considering two different procedures: high centrality
attacks, when a fraction of nodes or edges is deleted simultaneously
according to a list previously ordered by a chosen centrality index
(which is not unique and measures the structural importance of nodes
and edges in keeping the network cohesive) and high centrality
adaptive attacks when we attack individual components of the network
in accordance with a list iteratively ordered by a centrality index
and updated after each removal~\cite{netscibook}.  To test the
network's structural fragility we disrupt the network by node-based
High Degree Adaptive (HDA), High Betweenness Adaptive (HBA), High
Degree (HD), High Betweenness (HB) and Module-Based (MBA) attacks and
edge-based High Betweenness Adaptive (HBA), High Betweenness (HB),
Module-Based attacks (MBA) and Collective Influence (CI) (see
Fig.~\ref{sigmas}). The degree centrality is just the number of
connections a node has and the betweenness centrality basically
measures the fraction of shortest paths passing through a given
vertex~\cite{Iyer2013}.  The collective influence of a node takes into
account the degree of its neighbors at a given distance $l$ from it in
the following way
\begin{equation}
CI_{k}(i)=(k_{i}-1)\sum_{j\in \partial Ball(i,l)}(k_{j}-1)
\end{equation}
where $k_{i}$ is the node's degree and the $\partial Ball(i,l)$ is the
set of all nodes at a distance $l$ from node $i$. This method was
proven to be very close to the minimum dismantle
set~\cite{Morone:2015aa}.  The Module-Based
attack~\cite{Requiao-da-Cunha:2015aa} is related to the modular nature
of real networks, \textit{i.e.} the feature that complex networks tend
to group into clusters densely connected internally but only weakly
connected among them, those clusters are also called modules or
communities. The density of links connecting different communities
when compared to the internal density of edges is usually measured by
the network's modularity, $Q$ which ranges from $-1$ to $1$, and
depends slightly on the community extraction algorithm
used~\cite{gnbench}. It was previously shown that highly modular
networks are fragile to MBA~\cite{Requiao-da-Cunha:2015aa}. In this
sense, one would expect organized crime to show highly modular
features since the weak connection among communities would favor the
network's obscurity while the high internal density of communities
would make a proper scenario to efficiently run business
internally. Indeed, the network has a very high modularity either
using Louvain~\cite{blondel2008fast} ($Q=0.96$) or using
Infomap~\cite{Rosvall2009} ($Q=0.88$) methods.

To quantify the effects of each disruption strategy on the network we
measure the size of the largest connected component relative to the
network's original size, $G(q)$, as a function of the fraction of
removed objects, $q$. As pointed out in \cite{Perfo} the generalized
robustness of a network to a given attack strategy is given by the
metric:
\begin{equation}\label{robustness}
R=\frac{1}{N(1-G_{min})}\sum_{q=0}^{q_{max}}G(q)
\end{equation}
\newpage
where $N$ is the number of nodes in the network, $q_{max}$ is the
point at which the attack ends and $G_{min}$ is the value of the
relative size of the largest connected component at $q_{max}$.
Nonetheless, in order to evaluate the effectiveness of each strategy,
it is also important to measure the trade-off between robustness ($R$)
and the time ($t$) needed to compute the attack list. In this sense,
the performance of an attack is measured by the relation
$\mathcal{P}=t^{-1}\times R^{-1}$
where $t$ is the time taken to complete the procedure in seconds and
$R$ is the robustness~\cite{Perfo}.

\begin{figure}
\begin{center}
\includegraphics[width=0.95\columnwidth]{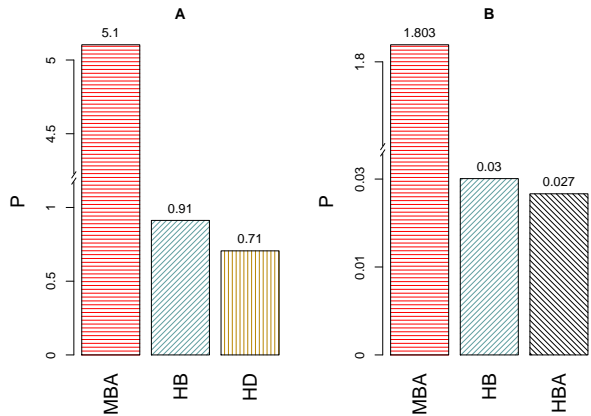}
\caption{The histograms show the performance of the three best attacks
  on the federal crimes network. Panel (A) shows Module-Based (MBA -
  horizontal red shades), High Betweenness (HB - inclined blue shades)
  and High Degree (HD - vertical golden shades) strategies for node
  removal, while panel (B) depicts MBA, HB and High Betweenness
  Adaptive (HBA - inclined black shades) attacks for edge-based
  attacks.\label{perf}}
\end{center}
\end{figure}

In accordance with these considerations, the attack strategy with
highest performance (see Fig.~\ref{perf}) is MBA both for node and edge
attacks, as expected for the network's high modularity. However, the
network is a little less robust to HBA, which in turn takes much more
time to compute. In other words, the network would be fully atomized
after removing approximately $2\%$ of its vertices and almost $5\%$ of
its edges by HBA. Besides, the deactivation point at which all
communities are detached from the core of the original graph is
reached by the MBA prescription when nearly $2\%$ of its edges or
nodes are removed. These results mean that even though node attacks
are in general more efficient than edge attacks, particularly in this
network both strategies are very similar-- for instance the edge MBA
has higher performance and similar robustness than the node HBA. This
is another important result, since the network would fragment
completely by traditional law enforcement actions (random attacks)
after the random failure of $80\%$ of nodes and $86\%$ of
edges. Another important feature is that the system is much weaker to
HBA and MBA attacks than to the novel CI strategy as depicted in
Fig.~\ref{sigmas}

The assortativity $(A)$ is another important aspect of the network.
In assortative networks $(A>0)$ nodes tend to connect to others with
similar degree and in dissortative networks $(A<0)$ high degree
vertices tend to attach to low degree nodes.  For the federal crimes
system, $A=0.02$. This is a well know phenomenon in social
networks~\cite{Newman:2002aa, Newman_mixing:2003}, \emph{ie} highly
prosperous people prefer to relate to people in the same social layer,
in business relationships entrepreneurs prefer to collaborate with
other big names in search for success, reputation, influence and
social status. Apparently, the same goes with criminal networks which
loosely is just a particular case of a business network.

A dynamic system is said to be controllable if one can get it to
evolve from any initial state to an arbitrary final state in a finite
time by an appropriate choice of external inputs. However, for very
large systems such as real networks it is more suitable to search for
a minimum subset of nodes whose control guarantees mathematically
control of the whole system. For instance, a dynamic variable such as
opinion, wealth or general tendency to commit a crime evolving in time
constricted to the criminal network's topology is reducible to a
minimum or zeroth level at least in principle if the system is
controllable. Recently, Liu \emph{et al}~\cite{Liu:2011aa} developed
the so-called structural controllability theory of directed networks,
which consists into identifying a minimum set of individual driver
nodes to achieve full control of complex networks, which was shown to
be equivalent to problem of maximum matching.  More recently, it was
shown that structural controllability can be achieved with a single
time-varying input suggesting that nodal dynamics is the key factor in
determining network controllability~\cite{Cowan:2012aa}.  Nonetheless,
the proposal is restricted to directed networks, which is not the case
studied here. Therefore, the exact controllability theory by Yuan
\textit{et al}~\cite{Yuan:2013aa} is more suited to this criminal
system. This framework is based on using the maximum geometric
multiplicity of the adjacency matrix to find the minimum set of
drivers required to fully control the system.  In this sense, consider
a linear system described by the following set of ordinary
differential equations:
\begin{equation}
\dot{\textbf{x}}=A\textbf{x}+B\textbf{u},
\end{equation}
where the vector \textbf{x} stands for the states of the nodes, $A$ is
the adjacency matrix of the network whose elements are $a_{ij}=1$ if
nodes $i$ and $j$ are connected and $a_{ij}=0$ otherwise, \textbf{u}
is the vector of controllers and $B$ is the control matrix. This
system is said to be controllable according to the framework proposed
by Yuan \textit{et al}~\cite{Yuan:2013aa} if we control a minimum
fraction of nodes (called drivers or controllers) given by:
\begin{equation}\label{drivers}
n_{D}=\frac{1}{N}max\{1,N-rank(A)\},
\end{equation}
In their seminal article Liu \textit{et al} have shown that,
counter-intuitively, social networks usually have very low $n_{D}$
values if compared to biological or infrastructure
networks. Confirming those results, the criminal network studied here
has a fraction of drivers $n_{D}=0.21$, suggesting that the whole
criminal system could be, in principle, controlled by only $2,076$
individuals.
%
%

\section{General Discussion \& Conclusion}
Thanks to a recent network data acquisition program by the Brazilian
Federal Police, we where able to study the network structure,
robustness and control of a large and unique criminal network covering
different classes of federal crimes all over Brazil. The system was
built directly by Federal Agents assigned to each investigation and
consists of 23,666 individuals in 35,913 undirected and unweighted
relationships. Surprisingly, the network has a giant component holding
more than $40\%$ of the nodes and $54\%$ of the edges.  We have showed
that the network shows small-world and scale-free behaviors being
``darker'' than traditional social networks, \textit{i.e.} combining
both low edge density values and low network efficiency.
 
The network is particularly weak to high betweenness adaptive attacks
and module-based attacks due mostly to its high modular nature. The
MBA attacks show a higher performance, meaning the deactivation point
where all communities are disjointed is reached at the expend of less
computational effort than the fully atomized phase reached by
centrality-based attacks with both critical points being not far from
each other.

Counter-intuitively, the network is highly controllable in the sense
that it is possible, in principle, to take any dynamical variable
(like opinion, wealth or criminal tendency) from its initial state to
arbitrary final states by controlling approximately only $20\%$ of its
nodes, in what appears to be a typical behavior of social
networks. However, even though the mathematical controllability of
this criminal system is guaranteed by the control of less than a
quarter of its nodes, it is not clear what this means in practical
terms for social systems. For instance, one is usually interested in
finding a desired stable final state or else the system will easily
move away, therefore mathematical controllability \textit{per se} does
not provide fully useful results. Besides, in social networks the
drivers are people and even the task of engineering a single input
becomes unclear and arguable both ethically and juridically. Thence a
deep understanding of social control is still a very open subject.

We argue that traditional imprisonment is equivalent to edge-based
attacks, while node-based attacks are more related to the death or
complete re-socialization of criminals.  However, although in general
it is more efficient to remove nodes than edges, particularly in this
network both strategies have similar results. Besides that, the
criminal data-bank is continuously growing (at the moment we submitted
this paper the size of the dataset reached more than 100,000 nodes)
making it virtually impossible to generate iteratively $N$
betweenness-based attack lists, which grows at best as $(N\times
E)^2$. Therefore, the best strategy to reduce criminality levels in
accordance with the topology of federal crimes in Brazil would be
either by educational or by prison polices with the targets chosen by
a modular approach depending on the political and practical
feasibility of each strategy-- for instance the prison system would
have to really work in cutting prisoners' social ties and in
re-socializing or educating them, which are \emph{de facto} very hard
tasks.

Our future research will focus experimentally on criminal networks
dynamics and effective control. We hope our results will help change
the approach of law enforcement agencies worldwide and thence lowering
crime level specially in Brazil. The authors would like to thank the
Institutional Defense Unit and the Organized Crime Fighting Unit of
the Brazilian Federal Police at Rio Grande do Sul.


\end{document}